\documentclass{article}
\usepackage{amsmath}

\newtheorem{theorem}{Theorem}
\newtheorem{acknowledgement}[theorem]{Acknowledgement}

\input{tcilatex}

\begin{document}

\title{Field tuned atom-atom entanglement via diople-dipole interaction}
\author{Yan-Qing Guo\thanks{%
E-mail: guoyqthp@yahoo.com}, Hai-Jing Cao, and He-Shan Song\thanks{%
Corresponding author: hssong@dlut.edu.cn} \\
%EndAName
Theor. Phys. Group, Dept. of Phys., Dalian Univ. of Tech.\\
Dalian, 116024, P. R. China}
\maketitle

\begin{abstract}
We propose a simple scheme, in which only one atom couples to a cavity
field, to entangle two two-level atoms. We connect two atoms with
dipole-dipole interaction since one of them can move around the cavity. The
results show that the peak entanglment does not depend on dipole-dipole
interaction strength but on field density at a certain controlling time. So
the field density can act as a switch for maximum entanglement (ME)
generation.

\end{abstract}

\baselineskip=15.5pt

\section{Introduction}

Recently, generation of entanglement in atomic systems has been intensely
paid attention to because of the motivations in the potential applications
in quantum information\cite{1,2,3} and computation processing\cite{4,5}. The
realization of easily controllable atomic entangled states turns out to be
one of the crucial and challenging tasks. A number of shcemes of generating
entanglement between atoms have been put forward to realizing quantum
teleportation\cite{6,7,8} and swapping\cite{9,10}. In many of the cases,
atoms are trapped in cavities or potential wells so that they can be
connected through directly exchanging real photons and then they can be
strongly entangled. While, practical applications in quantum information
processing require \textit{engineering} entangled atoms\cite{11}. This
expects operable atoms so that they can be moved to distance without losing
of information. To overcome this obstacle, many schemes have been proposed%
\cite{12,13,14,15}, for example, in Ref. [12], the atoms are assumed to move
in and out of the optical cavities and precisely received by detectors,
maximum entanglement (ME) are acquired by fixing controlling time. In this
paper, we propose a simple scheme to realize an easily engineered two-atom
entangled state. The advantage of this scheme is only one atom is trapped in
a cavity, and the other one can be spatially moved freely outside the
cavity. Further more, the amount of the entanglement can be manually tuned
by adjusting some key observable variables of the system.

\section{Model Description}

We consider a system constituted by two two-level identical atoms (1, 2) and
a single mode electromagnetic cavity field, as is shown in Fig. 1. Atom 1
resonantly interacting with the cavity field (sinelike line in the diagram)
is trapped in a slender microcavity. Atom 2 lies beside atom 1 out of the
cavity. We assume the cavity is so much narrow that atom 2 is close to atom
1. If the relative distance of two atoms and the de Broglie wavelength of
two atoms can compare, the dipole-dipole interaction (dashed line connecting
two atoms in the diagram) should be included. The whole device is located in
vacuum. 

The Hamiltonian of this system can be written as%
\begin{equation}
H=\hbar \omega (\sigma _{1}^{z}+\sigma _{2}^{z})+\hbar \omega
a^{+}a+g(a\sigma _{1}^{+}+h.c.)+\Gamma (\sigma _{1}^{+}\sigma _{2}^{-}+h.c)%
\text{.}
\end{equation}%
where $\sigma _{1}^{z}$, $\sigma _{2}^{z}$, $\sigma _{1}^{\pm }$ and $\sigma
_{2}^{\pm }$ are spin operators and raising (lowering) operators of atom $1$
and $2$ respectively, $a^{+}$, ($a$) is the creation (annihilation) operator
of field, $g$ is the coupling strength of atom 1 to field, while $\Gamma $
is the coefficient of atom-atom dipole-dipole interaction which has been
considered by many authors\cite{16,17}. In the invariant sub-space of the
global system, we can choose a set of complete basis of atom-field system as 
$\left| g,g,n+2\right\rangle $, $\left| e,g,n+1\right\rangle $, $\left|
g,e,n+1\right\rangle $, $\left| e,e,n\right\rangle $. On this basis, the
Hamiltonian of the system can be written as%
\begin{equation}
H=\left( 
\begin{array}{cccc}
(n+1)\omega & g\sqrt{n+2} & 0 & 0 \\ 
g\sqrt{n+2} & (n+1)\omega & \Gamma & 0 \\ 
0 & \Gamma & (n+1)\omega & g\sqrt{n+1} \\ 
0 & 0 & g\sqrt{n+1} & (n+1)\omega%
\end{array}%
\right) \text{.}
\end{equation}

We can easily get four eigenvalues of this Hermite Hamiltonian as%
\begin{equation}
E_{12}=(n+1)\omega \pm A\text{, }E_{34}=(n+1)\omega \pm B\text{,}
\end{equation}%
where, $A=\sqrt{C+D}/2$, $B=\sqrt{C-D}/2$ with $C=4ng^{2}+6g^{2}+2\Gamma
^{2} $, $D=2\sqrt{4(n+1)g^{2}\Gamma ^{2}+(g^{2}+\Gamma ^{2})^{2}}$. Also,
the corresponding four eigenvectors are

\begin{equation}
\left| \phi _{i}\right\rangle =\xi _{i}[x_{i1}\left| g,g,n+2\right\rangle
+x_{i2}\left| e,g,n+1\right\rangle +x_{i3}\left| g,e,n+1\right\rangle
+x_{i4}\left| e,e,n\right\rangle ]
\end{equation}%
with $i=1,2,3,4$, $\xi _{i}=1/\sqrt{\left| x_{i1}\right| ^{2}+\left|
x_{i2}\right| ^{2}+\left| x_{i3}\right| ^{2}+\left| x_{i4}\right| ^{2}}$,
and $x_{ij}$ ($j=1,2,3,4$) satisfy $x_{i1}=1$, $x_{i2}=(-1)^{i+1}\frac{C}{(%
\sqrt{n+2}g)}$, $x_{i3}=\frac{[C^{2}-(n+2)g^{2}]}{\sqrt{n+2}g\Gamma }$, $%
x_{i4}=(-1)^{i+1}\frac{C[C^{2}-(n+2)g^{2}-\Gamma ^{2}]}{\sqrt{n+1}\sqrt{n+2}%
g^{2}\Gamma }$, where $C=A$ for odd $i$ and $C=B$ for even $i$.

For a given initial state $\psi (0)$ of system, we can obtain the evoluted
dressed state of system $\psi (t)$ which is controlled by an unitary
transformation $U=e^{-iHt/\hbar }$. Also, $\psi (t)$ can be expanded as a
superposition of eigenstates $\phi _{i}$%
\begin{equation}
\left| \psi (t)\right\rangle =\sum\limits_{i}C_{i}(t)\left| \phi
_{i}\right\rangle \text{.}
\end{equation}%
The coefficients $C_{i}(t)$ are determined by solving Schrodinger Eqution,
so that%
\begin{equation}
C_{i}(t)=C_{i}(0)e^{-iE_{i}t/\hbar }\text{.}
\end{equation}%
The reduced density matrix of two-atom is obtained by tracing over the field
variables of system density matrix $\rho (t)=\left| \psi (t)\right\rangle
\left\langle \psi (t)\right| $, that is $\rho _{atom}(t)=Tr_{f}\rho
(t)=\sum\limits_{n}\left\langle n\right| \rho (t)\left| n\right\rangle $.
The initial condition $C_{i}(0)$ is easily given for an initial system state 
$\left| \psi (0)\right\rangle =\left| \psi (0)\right\rangle _{atom}\otimes
\left| \psi (0)\right\rangle _{field}$.

In the next section, we will discuss the two-atom entanglement induced by
the dipole-dipole interaction between atoms.

\section{Two-atom Entanglement Nature Under Cavity Field}

Wootters Concurrence, which has been proved to be effective in measuring the
entanglement of two qubits, is defined as\cite{18}%
\begin{equation}
C(\rho )=\max \{0,\lambda _{1}-\lambda _{2}-\lambda _{3}-\lambda _{4}\}\text{%
,}
\end{equation}%
where $\lambda _{i}$ are four non-negative squre roots of the eigenvalues of
the non-hermitian matrix $\rho (\sigma _{y}\otimes \sigma _{y})\rho ^{\ast
}(\sigma _{y}\otimes \sigma _{y})$ in decreasing order. In dealing with this
model, the Concurrence is simply determined by several density matrix
elements since most of the off-diagonal elements are eliminated due to the
adiabatic evolution. We choose the initial system state is a separable pure
state as%
\begin{equation}
\left| \psi (0)\right\rangle =\left| g\right\rangle _{2}\otimes \left|
g\right\rangle _{1}\otimes \left| n_{0}\right\rangle \text{.}
\end{equation}

Then the evolution of system state is determined by a series of parameters
space $(n$, $g$, $\Gamma )$. Fig. 2 shows two-atom entanglement under
parameters space $(1$, $5.0$, $0.5)$ for solid line and $(1$, $5.0$, $0.1)$
for dashed line. 

Generally, the entanglement shows local peaks, all the peaks present to be
covered under series of wave packets. This is caused by the Rabi oscillation
of atoms, and in fact the synchronization difference between the coupling of
atom 1 to field and atom 1 to atom 2. The curves show that \textit{larger}
dipole-dipole interaction can improve the local peak of the entanglement but
take no effect on the width of the peak of the amount of entanglement.
Physically, dipole-dipole interaction $\Gamma $ can be enhanced by reducing
the relative distance between two atoms or increasing the dipole polar
moment for each atom\cite{19}.

Fig. 3 shows the influence of different coupling of atom 1 to cavity on the
two-atom entanglement under same dipole-dipole coupling strength.
Surprisingly, weak atom-field coupling amplifies the packets of the
entanglement so that the amount and the width of each peak of the
entanglement are both enlarged. Especially, the first packet of the
entanglement is amplified most evidently. Since the coupling of intro-cavity
atom and cavity depends on the relative position of atom, $\mathbf{r}(x,y,z)$%
, in the cavity, such that $g=g_{0}\sin (k_{0}z)\exp [-(x^{2}+y^{2})/\omega
_{0}^{2}]$ with $g_{0}$ the peak coupling rate, $k_{0}$ and $\omega _{0}$
the wave vector and width of the cavity mode\cite{20}. The ideal case would
be fixing the atom to keep $g$ a constant so that the controlling time of
peak entanglement can be precisely operated.

In Fig. 4, we depict the entanglement under different intro-cavity field
densities $n$. The alternating of field density does not seem to take
influence on the local peak of the amount of the entanglement. While, the
width of the packet is slightly broadened for larger $n$. To distinctly
represent the entanglement under this situation, we illustrate two-atom
entanglement versus initial field density and dipole-dipole interaction
strength in Fig. 5. In computing the results, the controlling time is
restricted to be $4$. Note that the MEcan never be reached when $\Gamma $ is
small for arbitrary $n$. Surprisingly, the local peak entanglement arises
for a series of fixed $n$ whatever the value of $\Gamma $ be, for example,
the first peak emerges at about $n=4$, the second at about $n=14$. This
suggests us the initial field density acts as a tuning switch that
discretely controls the generation of ME at any specific time whatever the
dipole-dipole interaction strength be. Certainly, appropriate $\Gamma $ is
optimal, since the MEarises at about $\Gamma =0.4$ when $t=4$ and $%
n=4,14,\cdots $, in this system.

\section{Conclusion}

We proposed a simple scheme to entangle two two-level atoms that can be
realized physically. The Hamiltonian for the system was diagonalized to
obtain the eigenvalues. For a given initial global state, the evolutive
state of the two-atom subsystem was found to be entangled. The amount of the
entanglement is presented to depend on the coupling strength $g$ of
intra-cavity atom to field and the atomic dipole-dipole interaction strength 
$\Gamma $. It was found that larger $\Gamma $ and smaller $g$ benefit the
quantity and quality of peak entanglement. When these two characters were
fixed, the entanglement would be determined by the cavity field density and
the controlling time. And for a fixed controlling time, the ME took place at
a series of discrete field density intervals. From this point of view, the
field density $n$ acts as a ME generation tuning switch. The advantage of
this scheme is the extra-cavity atom can be moved so that the relative
distance between two atoms can be controlled, then $\Gamma $ is
controllable. Also, field density $n$ can be alternated using high quality
laser jet. So, this device may be further developed into an engineered
atom-atom entangler.

\begin{acknowledgement}
This work is supported by NSF of China, under Grant No. 10347103, 10305002
and 60472017.
\end{acknowledgement}

\bigskip 

{\huge Figure Captions:}

Fig. 1: Schematic diagram for field modulated two-atom entanglement model.
One atom is trapped by a electromagnetic field in a microcavity, the other
is located not far away beside the former atom out of the cavity. Two atoms
can be connected through dipole-dipole interaction.

Fig. 2: The evolution of two-atom entanglement versus time under different
dipole-dipole coupling. Solid line for $\Gamma =0.5$, dashed line for $%
\Gamma =0.1$.

Fig. 3: The evolution of two-atom entanglement versus time under different
atom-field coupling. Solid line for $g=1$, dashed line for $g=0.5$.

Fig. 4: The evolution of two-atom entanglement versus time under different
initial field densities. Solid line for $n=5$, dashed line for $n=6$.

Fig. 5: The evolution of two-atom entanglement versus initial field density
and atomic dipole-dipole interaction strength. The controlling time is fixed
at $t=4$.

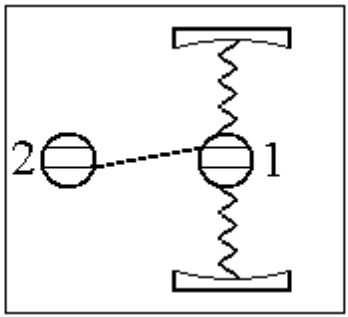
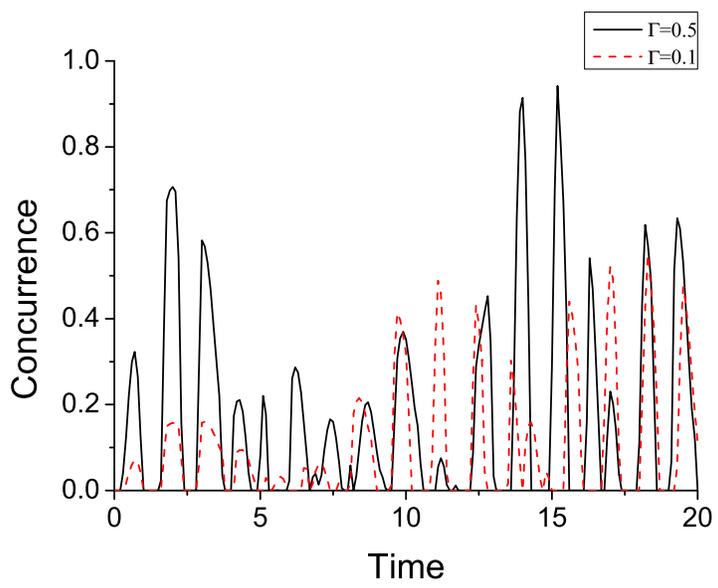
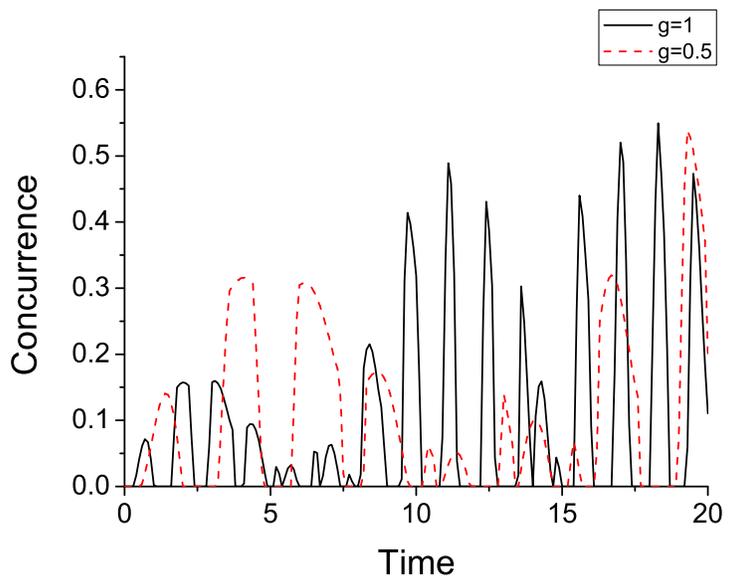
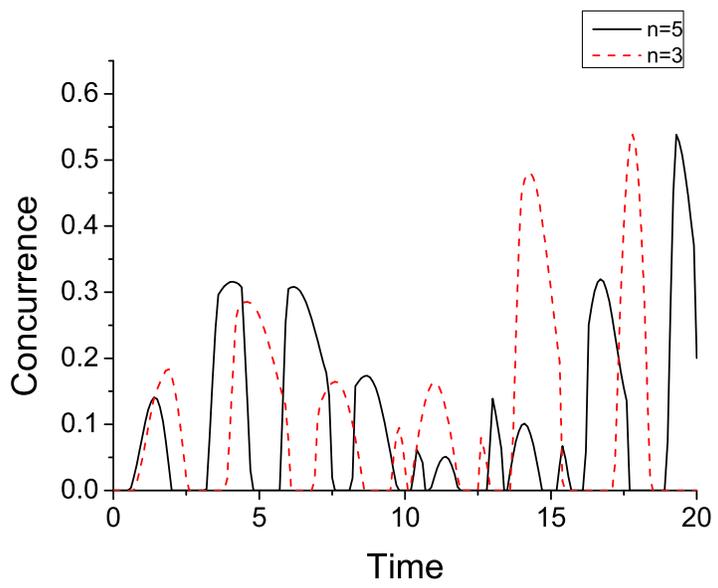
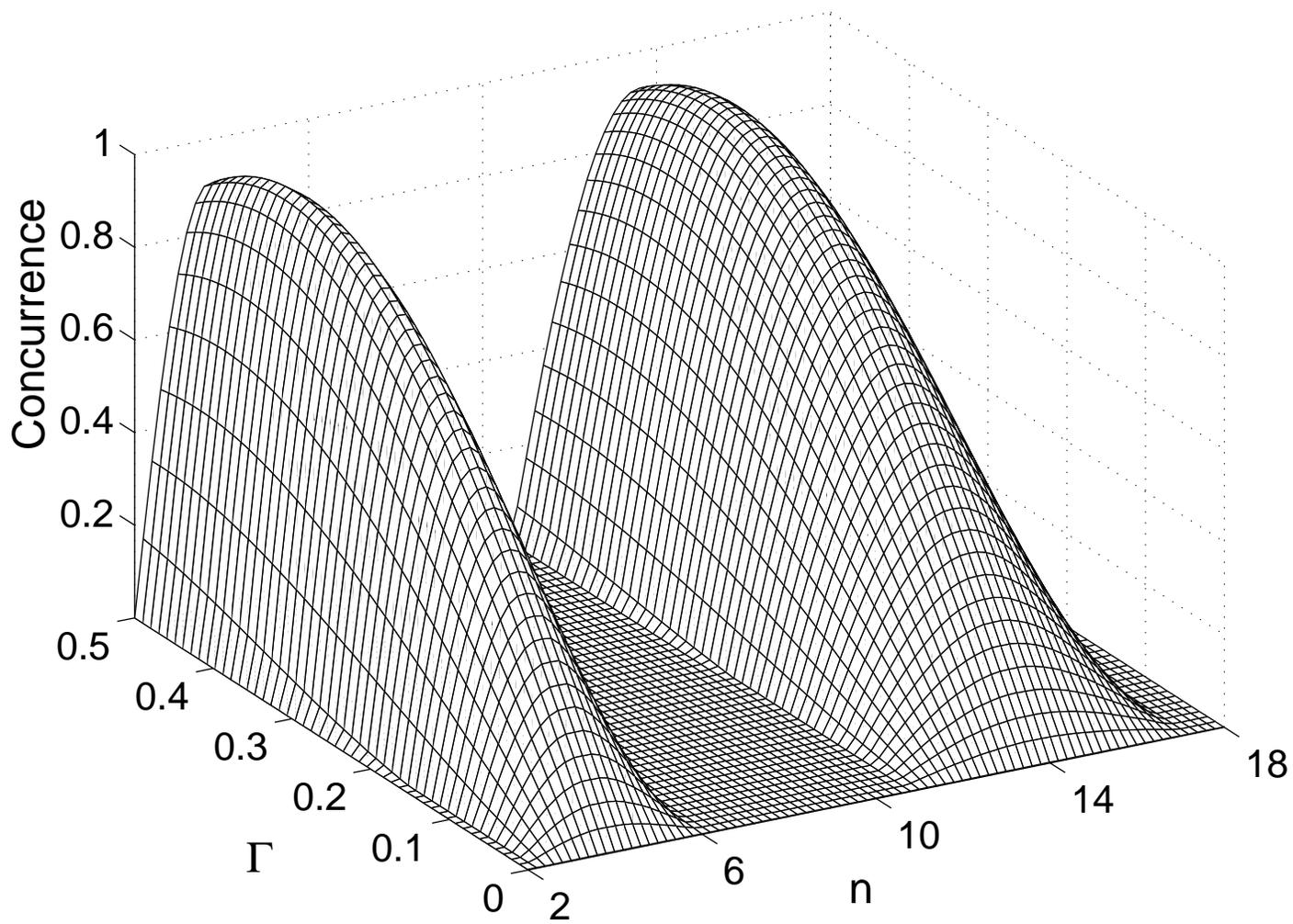
\end{document}